\documentclass[11pt,twoside]{article}

\usepackage{asp2006}
\usepackage{epsf}
\usepackage{psfig}
\usepackage{lscape}
\usepackage{graphicx}

\begin{document}
\title{Explaining the WMAP Haze with Neutralino Dark Matter}
\author{Gabriel Caceres}
\affil{Fermi National Accelerator Laboratory, Theoretical Astrophysics, Batavia, IL  60510}
\affil{Pennsylvania State University, Department of Astronomy and Astrophysics, University Park, PA 16802}

\begin{abstract}
It has been argued that the anomalous emission from the region around the Galactic Center observed by WMAP, known as the {\it WMAP Haze}, may be the synchrotron emission from relativistic electrons and positrons produced in dark matter annihilations. In particular, the angular distribution, spectrum, and intensity of the observed emission are consistent with the signal expected to result from a WIMP with an electroweak-scale mass and an annihilation cross section near the value predicted for a thermal relic. Here we revisit this signal within the context of supersymmetry and evaluate the parameter space of the Constrained Minimal Supersymmetric Standard Model. We find that over much of the supersymmetric parameter space the lightest neutralino is predicted to possess the properties required to generate the WMAP Haze. In particular, the focus point, $A$-funnel, and bulk regions typically predict a neutralino with a mass, annihilation cross section, and dominant annihilation modes that are within the range required to produce the observed features of the WMAP Haze. The stau-coannihilation region, in contrast, is disfavored as an explanation for the origin of this signal. If the WMAP Haze is indeed produced by annihilating neutralinos, prospects for future detection seem promising.
\end{abstract}

\section{Introduction}

The Wilkinson Microwave Anisotropy Probe (WMAP) has performed outstanding observations in order to determine cosmological parameters to greater precision~\citep{spergel}. While not its primary mission, WMAP can also study a number of astrophysical foregrounds, including synchrotron emission from supernova shock acceleration, emission from thermal and spinning dust, as well as ionized gas~\citep{Gold:2008kp}. Suprisingly, observations of the inner $20^{\circ}$ around the Galactic Center have revealed an excess microwave emission, known as the {\it WMAP Haze}, which does not appear to be a result of any of the standard foreground mechanisms~\citep{haze1a}.

Originally, the WMAP Haze was believed to be thermal bremsstrahlung emission from hot gas. However, the observed signal is inconsistent with a free-free spectrum in addition to the lack of an H$\alpha$ recombination line or X-ray emission~\citep{haze1b}. In light of this, the spectrum of the WMAP Haze leads one to interpret it as synchrotron emission with a very hard spectral index~\citep{haze1b}.

More recently, the WMAP Haze has been interpreted as the synchrotron emission from relativistic electrons and positrons produced in dark matter annihilations~\citep*{haze2}. In particular, it has been shown that dark matter in the form of weakly interactive massive particles (WIMPs) produced thermally in the early universe could naturally generate the observed emission. The angular distribution, spectrum and intensity of the WMAP Haze favor a WIMP with a mass in the range of 80 GeV to several TeV, a cusped halo profile within the several kiloparsecs around the Galactic Center ($\rho \propto r^{-1.2}$), and an annihilation cross section within a factor of a few of $3 \times 10^{-26}$ cm$^3$/s (the preferred value for a thermal relic)~\citep{haze2}. 

Many dark matter candidates have been proposed~\citep{review}, and among the best motivated are those which appear in supersymmetric extensions of the Standard Model. In particular, the lightest neutralino~\citep{neutralinodm1,neutralinodm2} is a very attractive and well studied dark matter candidate. Here, we revisit the dark matter interpretation of the WMAP Haze and study the supersymmetric parameter space which leads to a lightest neutralino with the properties required to generate the observed emission, summarizing the results of~\citet{hazesusy}. In Section 2 we briefly overview the results of~\citet{haze2}, where the dark matter properties required to generate the WMAP Haze were determined. In Section 3, we present the results of the supersymmetric parameter space scan and the implications for the WMAP Haze. In Section 4, we discuss and summarize the results.

\section{The Characteristics of Dark Matter Required To Generate The WMAP Haze}

Neutralinos (or other WIMP species) annihilating in the halo of the Milky Way produce a combination of gamma-rays, neutrinos, protons, antiprotons, electrons and positrons. The electrons and positrons which are produced move under the influence of the Galactic Magnetic Field, losing energy through inverse Compton scattering with starlight, emission from dust and the CMB, and through synchrotron emission. To determine the resulting electron/positron spectrum in the inner Galaxy, one solves the diffusion-loss equation~\citep{prop1,prop2}:
\begin{eqnarray}
\frac{\partial}{\partial t}\frac{dn_{e}}{dE_{e}} = \vec{\bigtriangledown} \cdot \bigg[K(E_{e},\vec{x})  \vec{\bigtriangledown} \frac{dn_{e}}{dE_{e}} \bigg]
+ \frac{\partial}{\partial E_{e}} \bigg[b(E_{e},\vec{x})\frac{dn_{e}}{dE_{e}}  \bigg] + Q(E_{e},\vec{x}),
\label{dif}
\end{eqnarray}
where $dn_{e}/dE_{e}$ is the number density of positrons per unit energy, $K(E_{e},\vec{x})$ is the diffusion constant, and $b(E_{e},\vec{x})$ is the rate of energy loss. The source term in the diffusion-loss equation, $Q(E_e, \vec{x})$, reflects both the distribution of dark matter in the Galaxy, and the mass, annihilation cross section, and dominant annihilation channels of the neutralino. Following \citet{haze2}, we adopt the following diffusion parameters: $K(E_e) \approx 10^{28} \, (E_{e} / 1 \, \rm{GeV})^{0.33} \,\rm{cm}^2 \, \rm{s}^{-1}$, and $b(E_e) = 5 \times 10^{-16} \, ({E_e} / 1 \, \rm{GeV})^2 \,\, \rm{s}^{-1}$. We also select boundary conditions corresponding to a slab of half-thickness 3 kiloparsecs, beyond which cosmic ray electrons/positrons are allowed to freely escape the Galactic Magnetic Field.

In order to calculate the spectral shape of the synchrotron signal, we adopt an average magnetic field strength of 10 $\mu$G within the inner few kiloparsecs of the Milky Way. To determine the fraction of the energy in electrons and positrons which is transferred into synchrotron emission, we compare the energy loss rates to synchrotron and inverse Compton scattering, which scale as the energy density in magnetic fields, $U_B$, and radiation fields, $U_{\rm rad}$, respectively. Thus the flux of synchrotron emission scales as $U_{B}/(U_B+U_{\rm rad})$. Although the true value could vary significantly, as our central estimate we adopt an average ratio of: $U_{B}/(U_B+U_{\rm rad}) = 0.25$, which is approximately the result of a magnetic field of 10 $\mu$G and a radiation field density of 5 eV/$\rm cm^3$. In order to account for this uncertainty, we will allow in our analysis values of $U_{B}/(U_B+U_{\rm rad})$ which are within the range of $0.1-1.0$.

\begin{figure}[t]
\centering\leavevmode
\mbox{
\includegraphics[width=3.5in,angle=0]{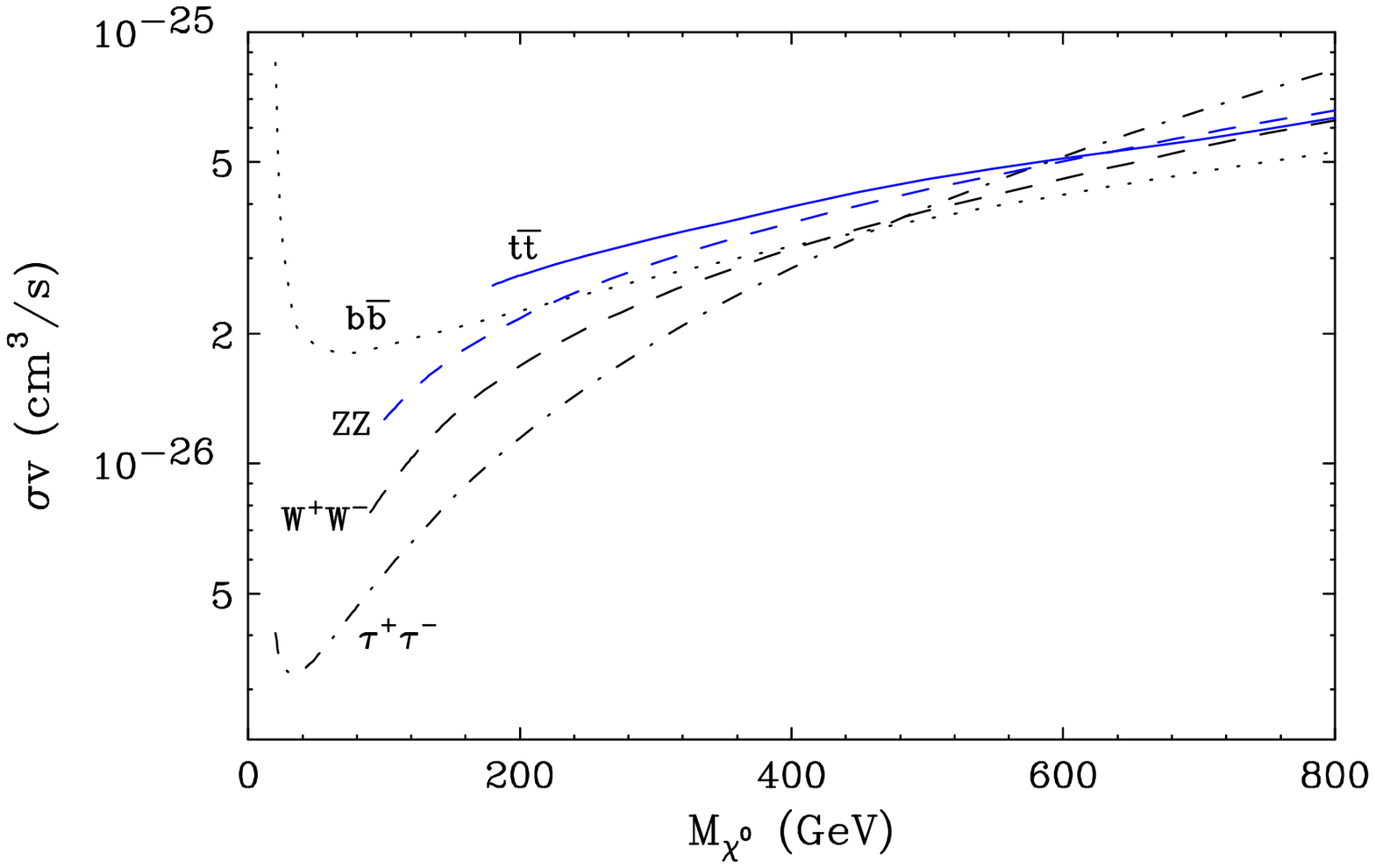}
}
\mbox{
\includegraphics[width=3.5in,angle=0]{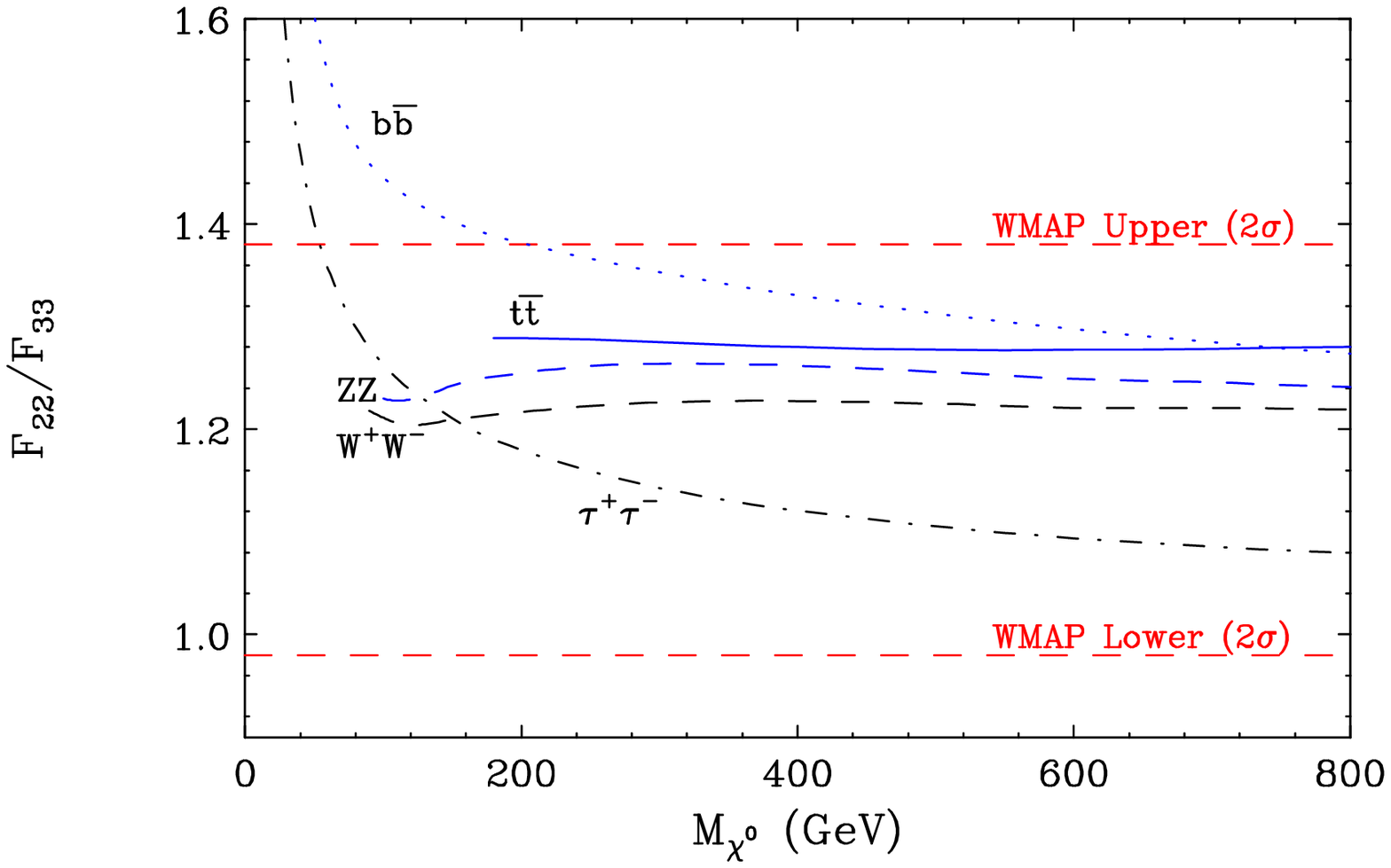}
}
\caption{{\it Left}: The neutralino annihilation cross section required to produce the observed intensity of the WMAP Haze as a function of mass, for several dominant annihilation channels. {\it Right}: The ratio of synchrotron intensities in WMAP's 22 GHz and 33 GHz frequency bands as a function of the neutralino's mass, for several dominant annihilation channels. Adapted from \citet{haze2}.}
\label{sigma}
\end{figure}

In \citet{haze2}, the WIMP mass and annihilation cross section required to generate the WMAP Haze were calculated for various dominant annihilation channels (after fixing the halo profile by matching to the angular distribution of the emission). In Fig.~\ref{sigma}, we plot this result, adapted slightly to the case of neutralino dark matter. In the top frame, the annihilation cross section (in the low velocity limit) required to normalize the synchrotron emission from dark matter annihilation products to the observed intensity of the WMAP Haze is shown, as a function of the neutralino mass, for several annihilation channels. From this plot, we see that an annihilation cross section of approximately $(1-5) \times 10^{-26}$ cm$^3$/s is required to produce the observed intensity of the WMAP Haze. This is remarkably similar to the value of approximately $3 \times 10^{-26}$ cm$^3$/s which is required of a thermal relic to be produced in the early universe with the measured dark matter abundance (in the absence of coannihilations, resonances or s-wave suppression). 

In the bottom frame of Fig.~\ref{sigma}, we turn our attention to the spectrum of the WMAP Haze. In particular, we show the ratio of synchrotron intensities produced in the 22 GHz and 33 GHz frequency bands of WMAP, again as a function of the neutralino's mass and dominant annihilation channel. The horizontal dashed lines are the (2$\sigma$) upper and lower limits of this ratio from measurements of the WMAP Haze~\citep{haze2}. From this frame, we see that for each of the annihilation channels shown, the resulting spectrum is consistent with that of the WMAP Haze, with the exception of light neutralinos ($m_{\chi^0} \la 200$ GeV) annihilating to b quarks (or a very light neutralino annihilating to tau leptons).

\section{The Constrained Minimal Supersymmetric Standard Model and the WMAP Haze}

The parameter space of the Constrained Minimal Supersymmetric Standard Model (CMSSM) consists of four continuous parameters: the universal scalar mass $m_0$, the universal gaugino mass $m_{1/2}$, the universal trilinear scalar coupling $A_0$, and the ratio of the vacuum expectation values of the two Higgs doublets $\tan \beta$, and one discrete parameter: the sign of the higgsino mass parameter $\mu$. Under the assumptions implicit in the CMSSM, the masses and couplings of the entire MSSM can be calculated from these five quantities. 

\begin{figure}[t]
\centering\leavevmode
\includegraphics[width=2.25in,angle=-90]{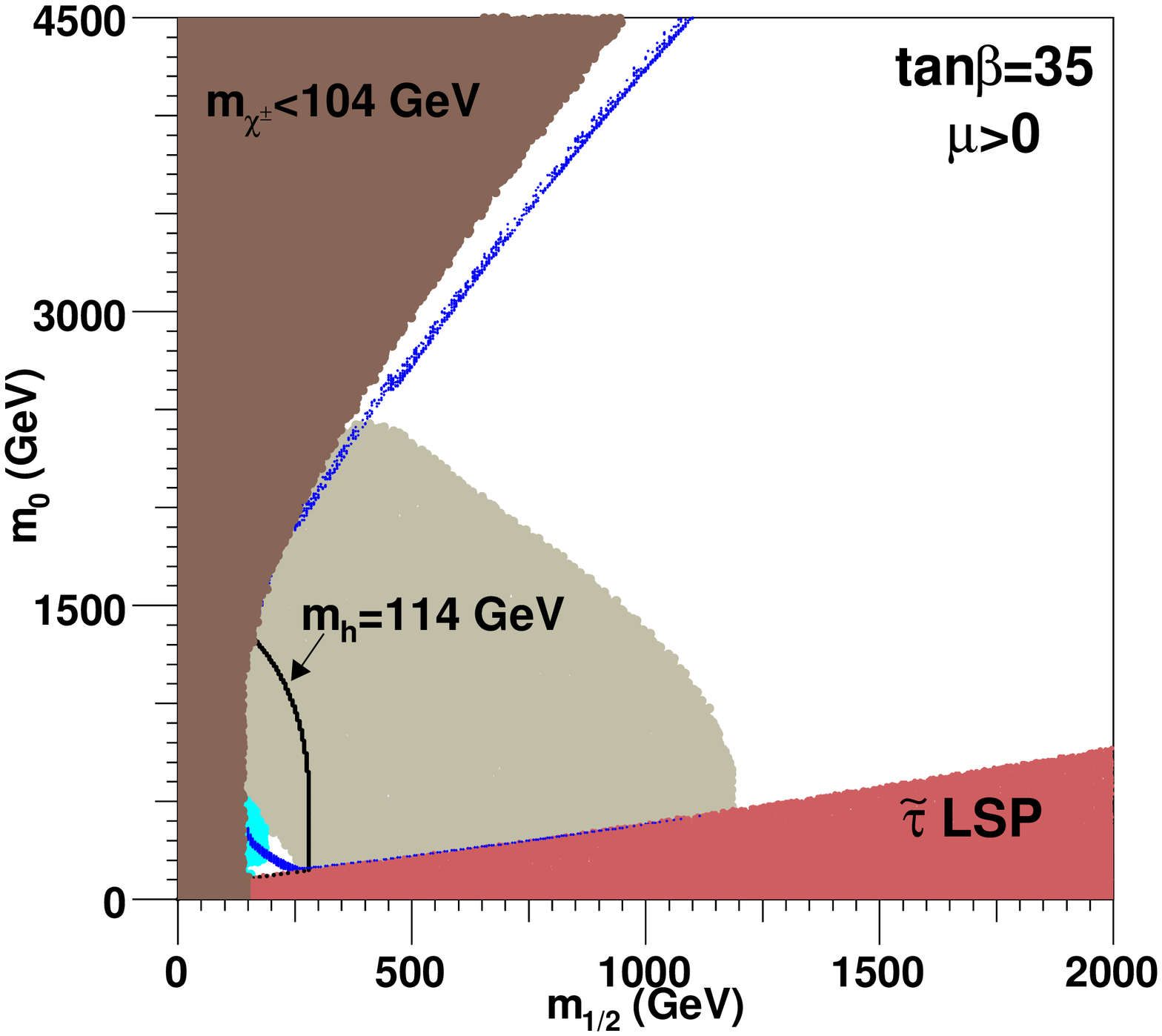}
\includegraphics[width=2.25in,angle=-90]{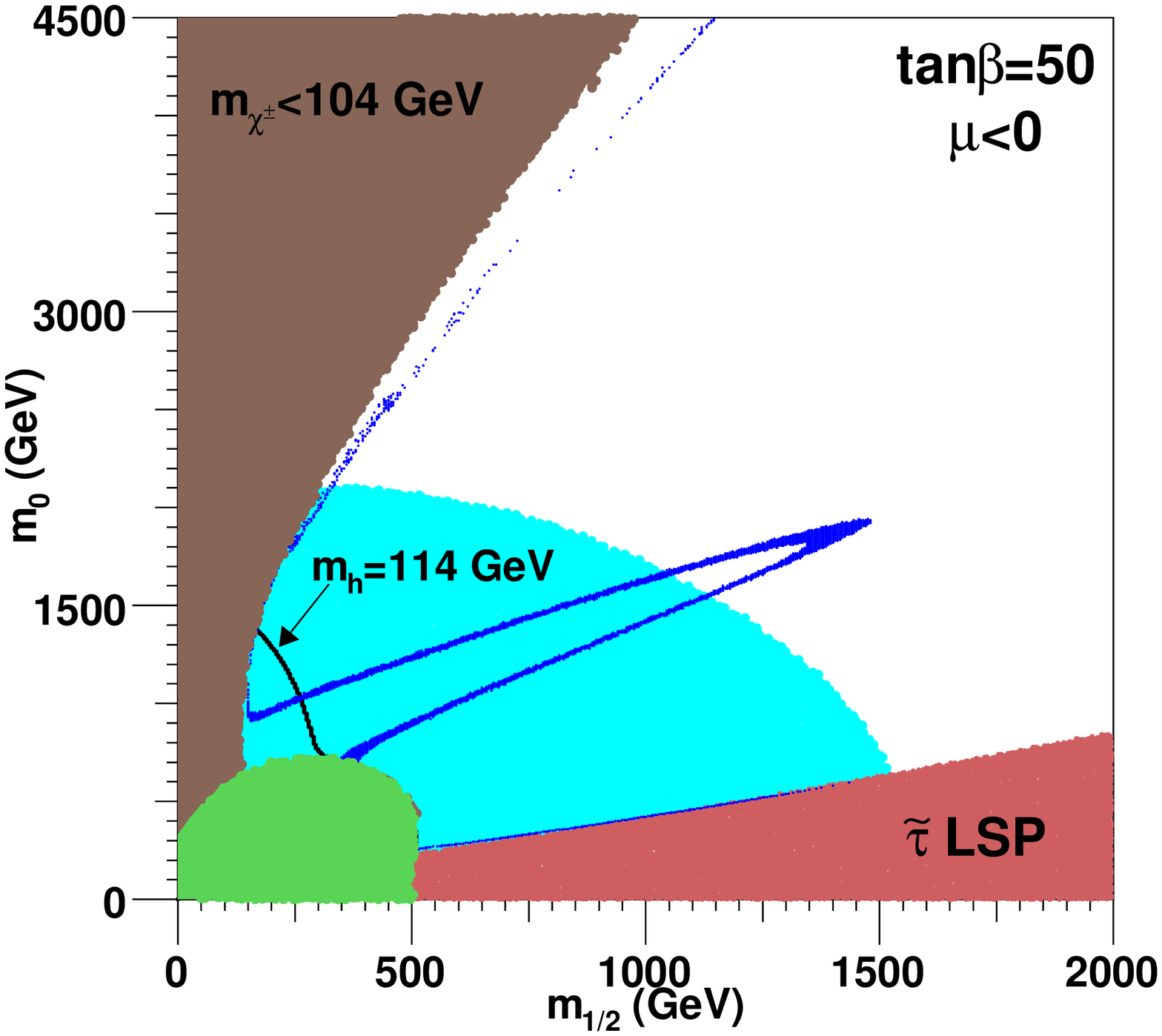}
\caption{Constraints on the CMSSM parameter space, including the region (blue) with an acceptable neutralino relic density, $0.0913 < \Omega_{\chi^0} h^2 < 0.1285$ \citep{spergel}. The shaded regions are disfavored/preferred by various bounds as explained in the text. In each frame, we have used $A_0=0$.}
\label{mzeromhalf}
\end{figure}

In Fig.~\ref{mzeromhalf}, we show a sample of the phenomenological features of the CMSSM parameter space. In each frame, the narrow blue region predicts a thermal abundance of neutralinos which is within the cold dark matter density range determined by WMAP ($0.0913 < \Omega_{\chi^0} h^2 < 0.1285$, using 3$\sigma$ errors)~\citep{spergel}. The upper left region of each frame is excluded by the LEP chargino bound ($m_{\chi^{\pm}}> 104$ GeV)~\citep{pdg}. In the lower right region, the lightest supersymmetric particle is a stau, and thus does not provide a viable dark matter candidate. Also shown are the contour corresponding to the LEP Higgs mass bound ($m_h > 114$ GeV)~\citep{pdg}, a lightly shaded regions corresponding to the parameter space preferred by measurements of the muon's magnetic moment  ($3.1\times 10^{-10} < {\delta a_{\mu}} < 55.9\times 10^{-10}$, using 3$\sigma$ errors)~\citep{gm2SM}, and colored cyan is the region disfavored (at the 2$\sigma$ level) by measurements of the $b->s\gamma$ branching fraction~\citep{bsgam}. Although we have not explicitly included constraints from $B_s->\mu^+\mu^-$, this is expected to be relevant only for the combination of very large values of $\tan \beta$ and small values of the pseudoscalar Higgs mass, which is not the case in any of our favored regions. We have used the DarkSUSY package to calculate these quantities~\citep{darksusy}.

While much of the parameter space shown in Fig.~\ref{mzeromhalf} appears to be ruled out or strongly disfavored by the experimental constraints imposed, we should note that departures from the assumptions implicit in the CMSSM can alter these restrictions. Thus we will continue to consider some regions of the parameter space which might otherwise be ruled out, although they will remain color-coded for the reader who prefers to disregard them. 

From Fig.~\ref{mzeromhalf} we see that only a small fraction of the parameter space predicts an abundance of neutralinos consistent with the measured density of dark matter. In particular, neutralino dark matter is expected to be overproduced relative to the observed dark matter abundance over the majority of the supersymmetric parameter space. The regions which do provide an acceptable dark matter abundance can be classified as follows:
\begin{itemize}
\item{{\it The focus point region}: For large values of $m_0$ and moderate or large values of $\tan \beta$, the lightest neutralino is a mixed bino-higgsino and, as a result, possesses large couplings which enable it to annihilate efficiently.}
\item{{\it The stau coannihilation region}: In the parameter space near the boundary of the $\tilde{\tau}$ LSP region, the lightest neutralino is nearly degenerate with the lightest stau. In these points, coannihilations between the lightest stau and lightest neutralino in the early universe lead to an acceptable density of neutralino dark matter.}
\item{{\it The bulk region}: The parameter space with light $m_0$ and light $m_{1/2}$ contains many light sparticles which, in some cases, enable the lightest neutralino to annihilate efficiently.}
\item{{\it The $A$-funnel region}: The parameter space with large $\tan \beta$ contains regions in which the lightest neutralino is able to annihilate efficiently through the CP-odd Higgs boson resonance $\chi^0 \chi^0 \rightarrow A \rightarrow f \bar{f}$.}
\end{itemize}

\begin{figure}[t]
\centering\leavevmode
\includegraphics[width=1.52in,angle=-90]{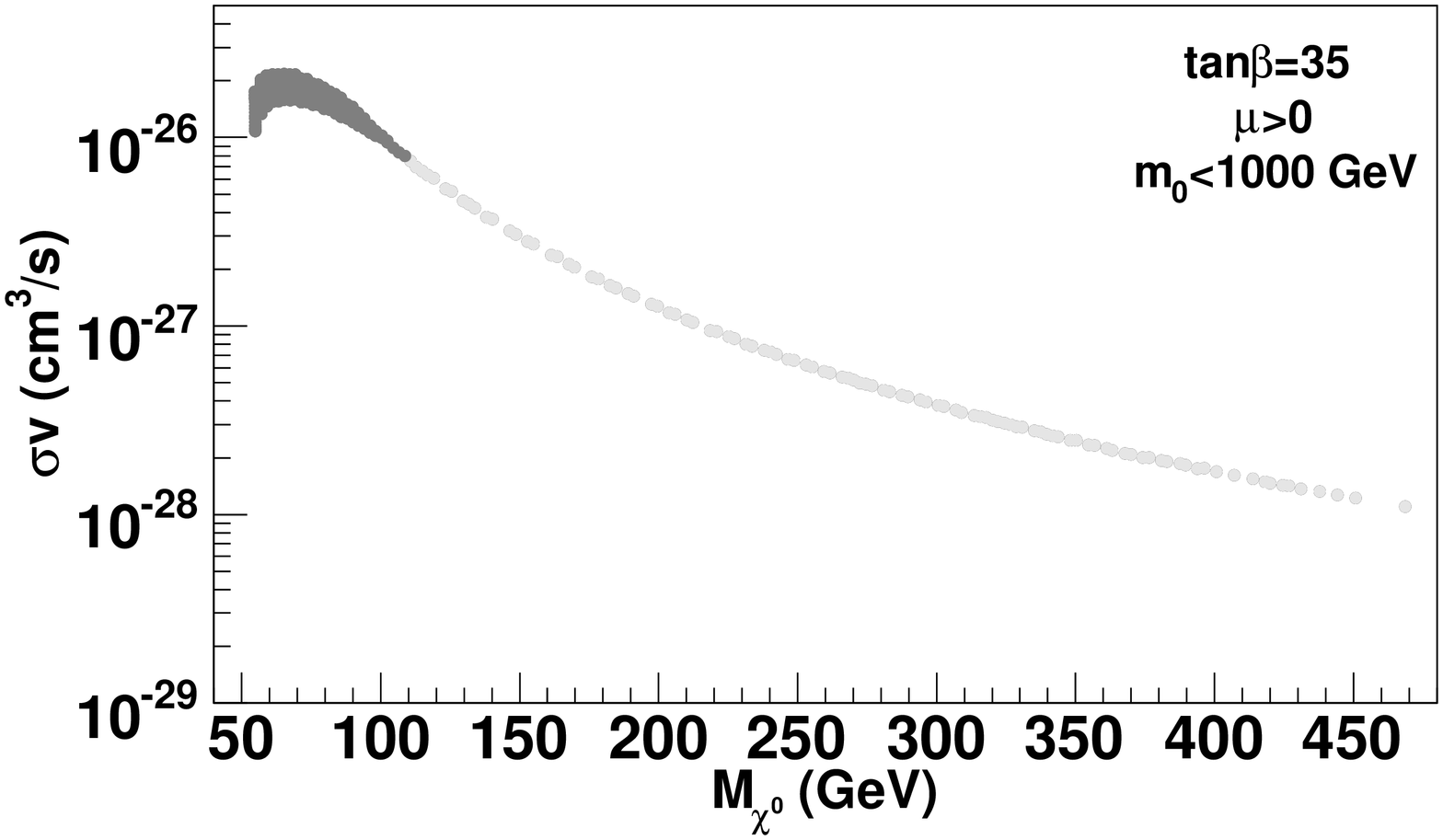}
\includegraphics[width=1.52in,angle=-90]{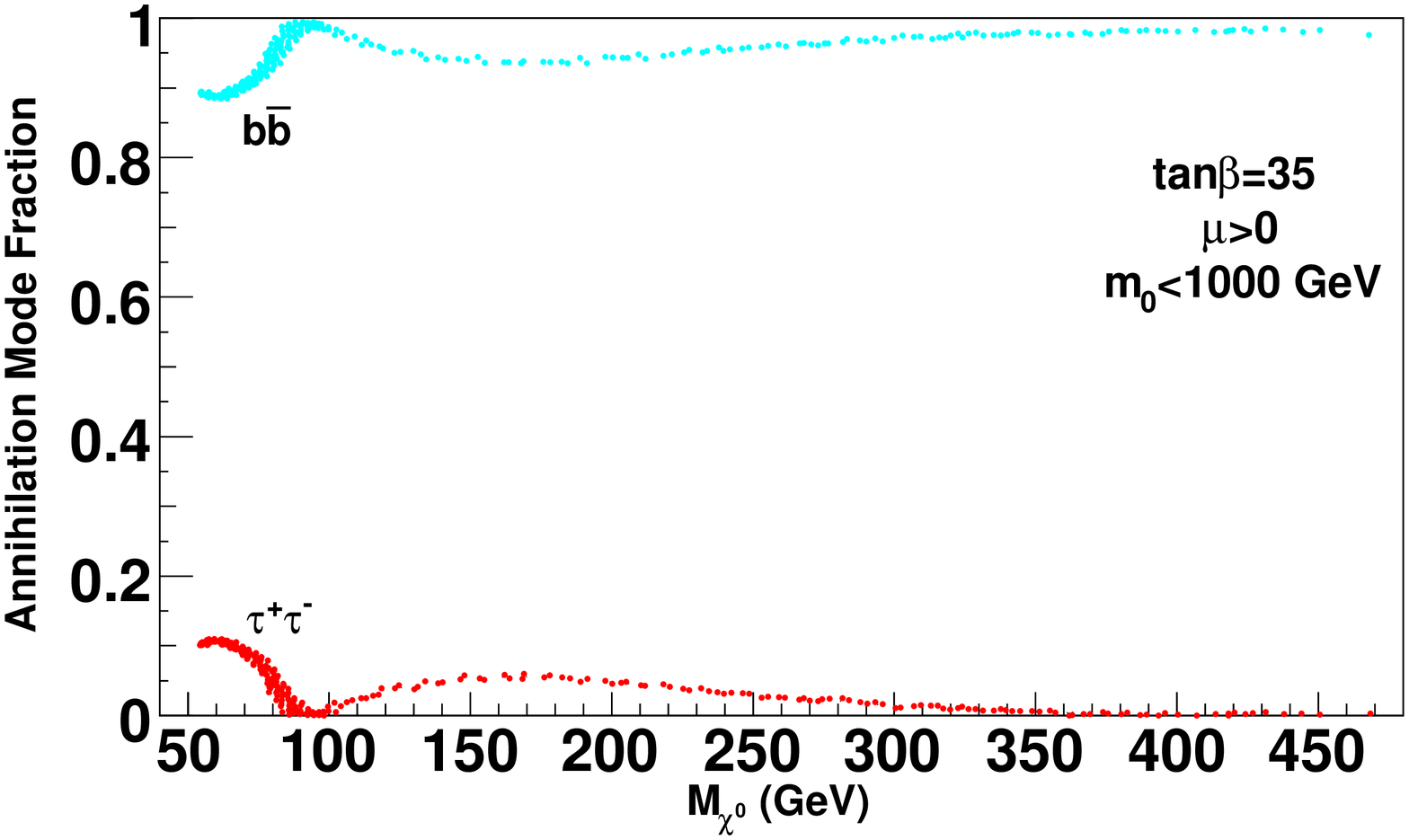}\\
\vspace{0.23cm}
\includegraphics[width=1.52in,angle=-90]{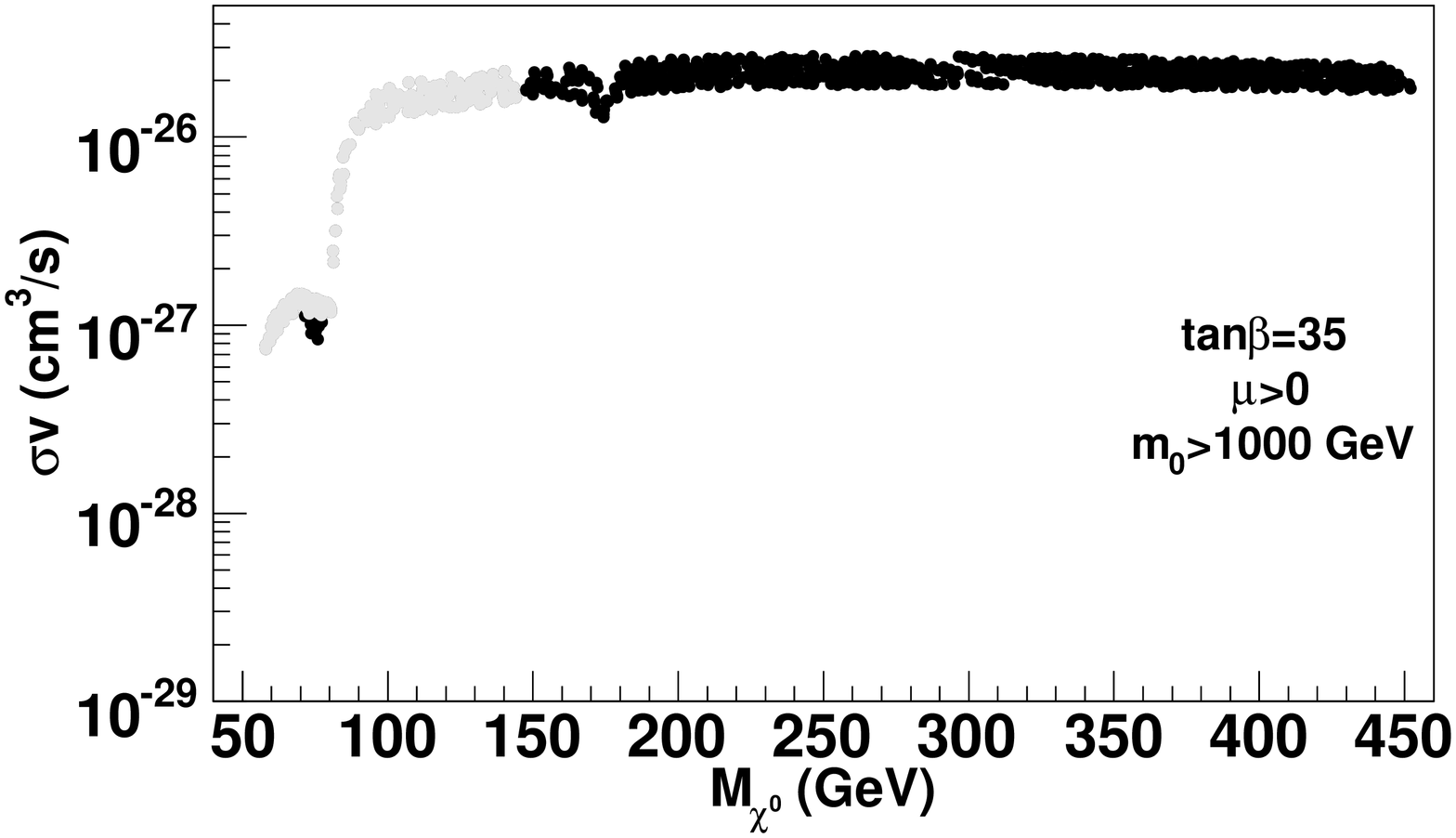}
\includegraphics[width=1.52in,angle=-90]{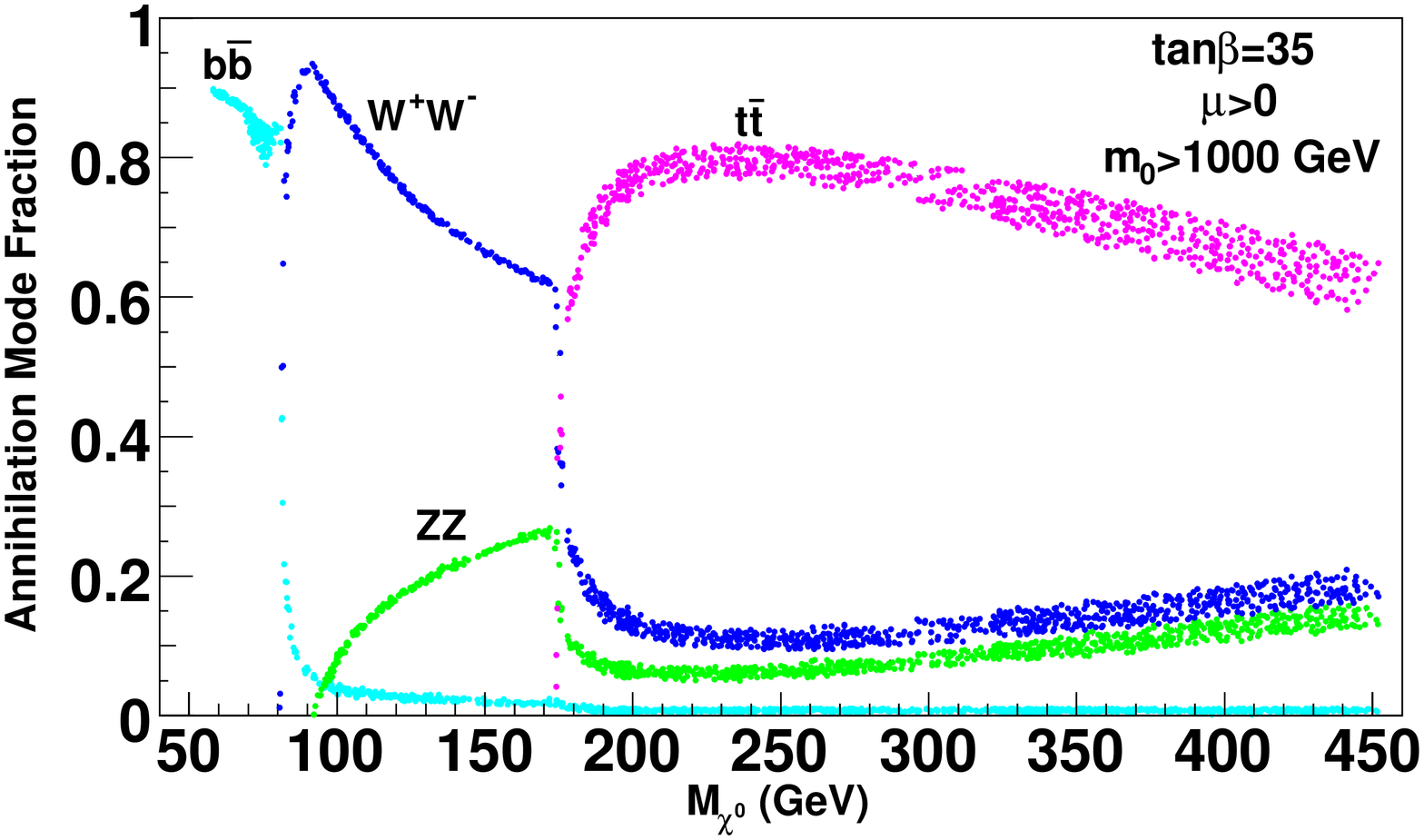}\\
\vspace{0.23cm}
\includegraphics[width=1.52in,angle=-90]{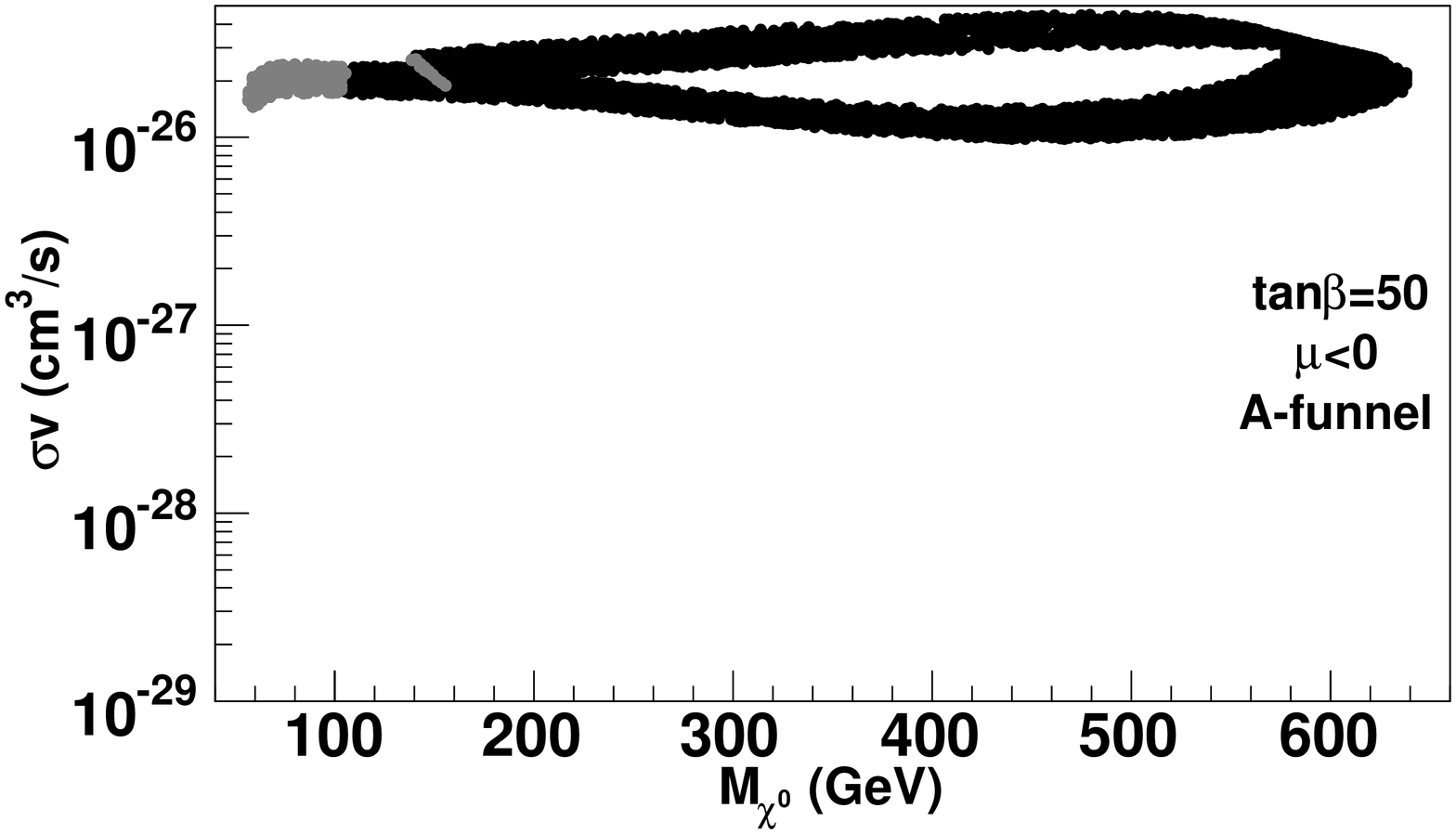}
\includegraphics[width=1.52in,angle=-90]{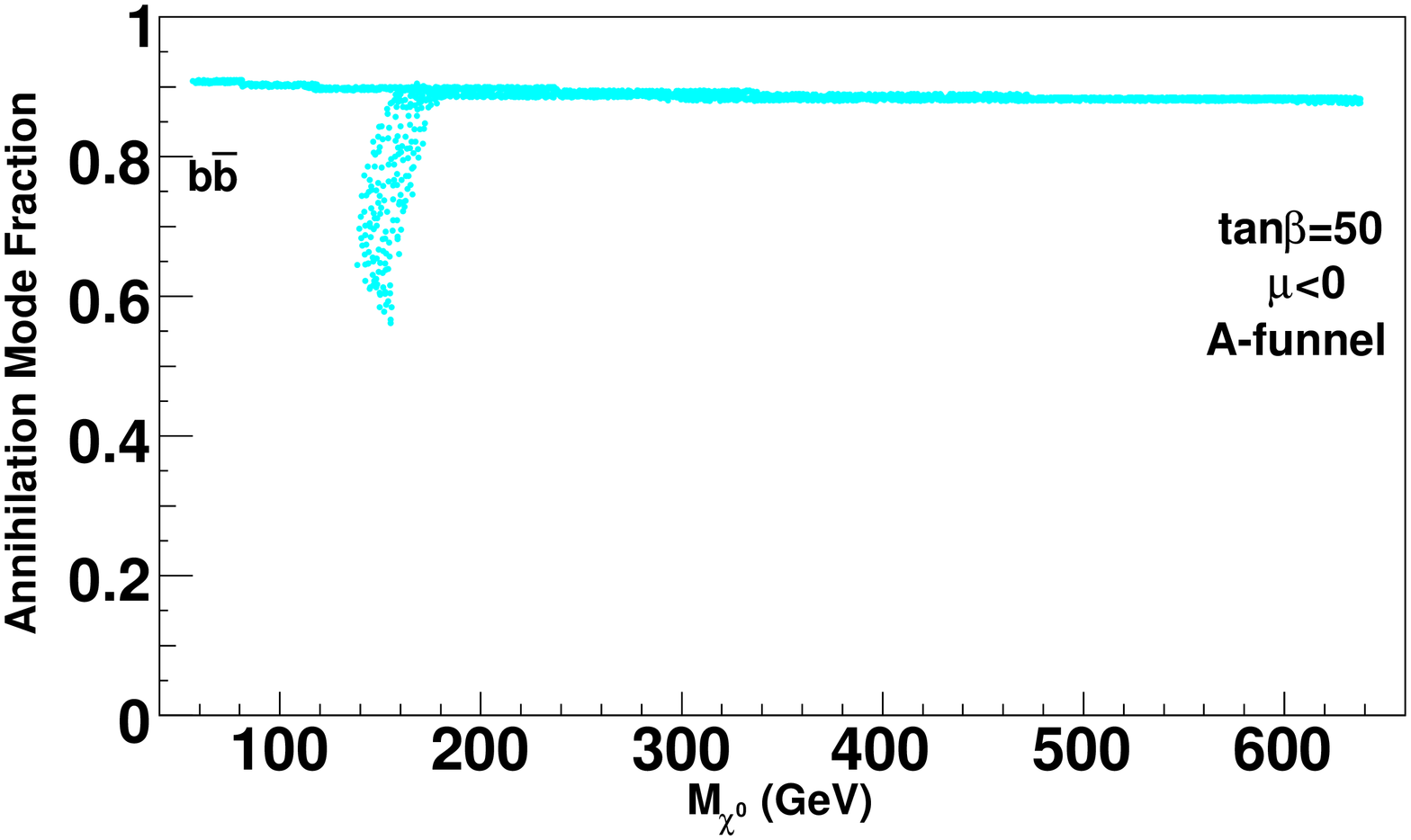}
\caption{{\it Left}: The low-velocity neutralino annihilation cross section. Shaded regions correspond to experimental bounds, see text for more details. {\it Right}: The fraction of neutralino annihilations which produce various Standard Model final states. We plot only the modes that contribute 12\% or more for at least one value of the neutralino mass in a given frame.}
\label{cs_mode}
\end{figure}

Focusing on the regions of parameter space which lead to an acceptable abundance of neutralino dark matter, in Fig.~\ref{cs_mode} we plot the neutralino annihilation cross section (left) and the branching fraction to various Standard Model final states (right), in the low-velocity limit as is appropriate for the WMAP Haze and indirect detection in general. For all frames, each point shown corresponds to a point in Fig.~\ref{mzeromhalf} that predicts a neutralino dark matter abundance consistent with the measured dark matter density, and that does not violate the LEP chargino mass bound. The dark gray shading on each left frame corresponds to regions disfavored by the LEP Higgs mass bound, while the light gray (black) regions are preferred (disfavored) by the measurements of $\delta a_\mu$ and $BF(b->s\gamma)$. 

The top two frames of Fig.~\ref{cs_mode}, labeled $m_0 < 1000$ GeV, correspond to the stau-coannihilation and bulk regions, which mostly annihilate to produce $b\bar{b}$ with a smaller contribution from $\tau^+ \tau^-$. The middle frames ($m_0 > 1000$ GeV) show the focus point region, where we see that neutralino annihilations produce mostly heavy quarks, gauge bosons, or a combination thereof. In the $A$-funnel region, the bottom frames of Fig.~\ref{cs_mode}, neutralino annihilations proceed largely to $b\bar{b}$.
 
Combining the characteristics required to explain the WMAP Haze, shown in Fig.~\ref{sigma}, with the properties of the neutralino, shown in Fig.~\ref{cs_mode}, we can determine the specific parts of the parameter space which may be able to explain the excess emission. Although the stau-coannihilation region predicts annihilation cross sections well below the range preferred in Fig.~\ref{sigma}, much of the focus point, bulk (with large $\tan \beta$), and $A$-funnel regions naturally predict a cross section very close to that required to produce the WMAP Haze. More precisely:
\begin{itemize}
\item{{\it The focus point region}: The entire focus point parameter space with $M_{\chi^0} > M_W$ naturally produces a signal in agreement with the spectrum and intensity of the WMAP Haze.}
\item{{\it The stau coannnihilation region}: Neutralinos in the stau-coannihilation region consistently under-produce the intensity of synchrotron emission relative to the intensity of the WMAP Haze.}
\item{{\it The bulk region}: Although in much of the bulk region the neutralino is light and annihilates largely to $b\bar{b}$, leading to a spectrum too soft to accommodate the WMAP Haze, we find that in the parameter space near $\tan \beta \sim 50$, $\mu >0$, and $M_{\chi^0} \sim 125-300$ GeV, the synchrotron emission is consistent with the observed properties of the WMAP Haze.}
\item{{\it The $A$-funnel region}:  Much like the focus point, the $A$-funnel region consistently predicts a neutralino annihilation cross section near the value required to normalize the WMAP Haze. So long as $M_{\chi^0} \ga 125$ GeV, $A$-funnel neutralinos are consistent with being the source of this signal due to a small admixture of $\tau^+\tau^-$(not shown in Fig.~\ref{cs_mode} because the contribution is less than 12\%).}
\end{itemize}

\section{Discussion and Conclusions}

It has previously been shown that the anomalous emission from the inner Milky Way known as the WMAP Haze can be generated by a WIMP with a mass within the range of 80 GeV to several TeV and an annihilation cross section near the value predicted for an $s$-wave annihilating thermal relic. Here, we have studied the possibility that annihilating neutralinos are the source of this signal. Confining our study to the Constrained Minimal Supersymmetric Standard Model (CMSSM), we find that a large fraction of the phenomenologically viable parameter space naturally leads to an annihilation cross section and spectrum of annihilation products consistent with the observed properties of the WMAP Haze (both the spectrum and intensity). In particular, the focus point, $A$-funnel, and high $\tan \beta$ bulk regions of the CMSSM parameter space are each well suited for generating this anomalous signal. We find that neutralinos in the stau coannihilation, or low $\tan \beta$ bulk region, in contrast, generate a spectrum of synchrotron emission which is either too faint, too soft, or both, to account for the WMAP Haze.

Although not discussed in the present work, if the WMAP Haze is in fact generated by annihilating neutralinos, then the prospects for direct and indirect detection are each promising. In particular, direct detection experiments are currently quite likely to be within 1-2 orders of magnitude of the sensitivity required to detect such a particle. The prospects for next generation neutrino telescopes such as IceCube are also quite encouraging, especially in the focus point region of the parameter space. Under the full constraints of the CMSSM, only the $\tan \beta \approx 50$, $\mu>0$, $m_0<1000$ GeV (bulk region) is consistent with the WMAP Haze yet beyond the reach of very near future experiments. 

We refer the interested reader to~\citet{hazesusy} for further details, including a more extensive discussion of the supersymmetric parameter space and the regions consistent with being the source of the WMAP Haze, as well as the implications for direct and indirect detection experiments.

\acknowledgements This work was carried out in collaboration with Dan Hooper. We would like to extend our gratitude to the SnowPAC 2009 organizing committee for their hospitality and work arranging this conference. We would also like to thank Greg Dobler and Doug Finkbeiner for very helpful discussions. This work has been supported by the US Department of Energy and by NASA grant NAG5-10842. GC has been supported by the Fermilab Summer Internships in Science and Technology Program.

\end{document}